\documentstyle[preprint,aps,floats,epsf]{revtex}

\def\Im{{\cal I}\!m\,}
\def\Re{{\cal R}\!e\,}
\def\bq{\begin{equation}}
\def\eq{\end{equation}}
\def\ba{\begin{eqnarray}}
\def\ea{\end{eqnarray}}

\begin{document}

\font\fortssbx=cmssbx10 scaled \magstep2
\hbox to \hsize{
\hbox{\fortssbx University of Wisconsin - Madison}
\hfill\vtop{\hbox{\bf MADPH-98-1024}
            \hbox{January 1998}} }

\vspace*{.5in}

\begin{center}
\vspace*{0.3in}
{\large\bf Unitarization of Gluon Exchange Amplitudes \\
and Rapidity Gaps at the Tevatron }\\
\vskip1cm
R.~Oeckl$^{1,2}$ and D.~Zeppenfeld$^1$\\
\vskip.5cm
$^1${\it Department of Physics, University of Wisconsin, Madison, 
WI 53706, USA}\\
$^2${\it DAMTP, University of Cambridge, Cambridge CB3 9EW, UK}
\end{center}

\vskip1in

\begin{abstract}
Rapidity gaps between two hard jets at the Tevatron have been interpreted
as being due to the exchange of two gluons which are in an overall
color-singlet state. We show that this simple picture involves unitarity
violating amplitudes. Unitarizing the gluon exchange amplitude leads to
qualitatively different predictions for the fraction of $t$-channel color
singlet exchange events in forward $qq$, $qg$ or $gg$ scattering, which 
better fit Tevatron data.
\end{abstract}

\thispagestyle{empty}
\newpage

\section{Introduction}

Over the past few years rapidity gaps, i.e. pseudorapidity regions without
hadronic activity, have been observed in hadronic collisions at both the
HERA $ep$ collider~\cite{HERAgap} and in $p\bar p$ collisions at the Fermilab
Tevatron~\cite{D0gap,CDFgap,D0gapnew,CDFgapnew}.
Such rapidity gaps are widely attributed to the exchange of color singlet 
quanta between incident partons~\cite{low,bjgap,others,rio}, the exchange 
of two gluons in a color singlet state being the simplest such 
model~\cite{low}. At the Tevatron, a fraction $f_{gap}\approx 1\%$
of all dijet events with jet transverse energies $E_{Tj}\agt 20$~GeV
and jet separations of more than three units of pseudorapidity
exhibit rapidity gaps between the jets. This observation is particularly
striking since it demonstrates that color singlet exchange effects in QCD
events are relevant at momentum transfers of order 1,000~GeV$^2$, raising
the hope that perturbative methods can be used for quantitative descriptions.

A gap fraction of order one percent was in fact predicted by
Bjorken~\cite{bjgap}, in terms of a fraction $f_s\approx 0.15$ of dijet events
which are due to $t$-channel color-singlet exchange and a survival probability
$P_S$ of rapidity gaps of order 10\%~\cite{bjgap,gotsman},
\bq
\label{eq:Ps}
f_{gap} = f_s\; P_S \; .
\eq
Here the survival probability estimates the fraction of hard dijet events
without an underlying event, i.e. without soft interactions between the
other partons in the scattering hadrons. Such multiple interactions
would fill the rapidity gap produced in the hard scattering process.
For $Q\bar q$ elastic scattering, Bjorken estimated the color-singlet
fraction $f_s$ in terms of the imaginary part of the two-gluon $t$-channel
exchange amplitude, which is
known to dominate the forward scattering amplitude for $t$-channel
color-singlet exchange. In impact parameter space, at impact parameters small
compared to $R={\cal O}(1/\Lambda)$, the result is
\ba
f_s^{\rm impact} & = &
{2\over 9}\; \left|{1\over 2}\;\alpha_s\left({1\over b^2}\right)\;
{\rm log}{R^2\over b^2}\right|^2
 =  {1\over 18}\; \left|{12\pi\over (33-2n_f)\;
{\rm log}{1\over b^2\Lambda^2} }\;{\rm log}{R^2\over b^2}\right|^2 \;
\nonumber \\
& \approx & {1\over 2}\; \left|{4\pi\over 33-2n_f}\right|^2 = 0.15\; .
\label{eq:fBjorken}
\ea
Here 2/9 is the relative color factor of the two-gluon color-singlet to the 
one-gluon color-octet exchange cross section and $R$ is an infrared cutoff
parameter which regularizes the two-gluon loop-integral. This model for the 
calulation of the color singlet fraction $f_s$, with the two gluon-exchange 
amplitude replaced by its imaginary part, will be called the two-gluon exchange
model in the following.

In this  model, the color singlet fraction grows with
the color charge of the scattered partons. For $qg$ and $gg$ elastic
scattering $f_s$ would be larger by factors $9/4$ and $(9/4)^2$,
respectively~\cite{rio}. This results in a substantial decrease of the 
observable gap fraction as the contribution from gluon induced
dijet events  is reduced, {\it e.g.} by increasing the average transverse
momentum of the observed jets and thereby the Feynman-$x$ values of the
incident partons. Such measurements have recently been reported by both the
CDF~\cite{CDFgapnew} and the D0~\cite{D0gapnew} Collaborations, and
no such effect is observed. In fact, the D0 data are compatible with a slight
increase of the gap fraction with increasing jet $E_T$, casting doubt on
the validity of the two-gluon exchange model~\cite{halzen}.

In this paper we reconsider the basic ideas behind the two-gluon exchange
model. We demonstrate its limitations and show that, even when starting
from this perturbative picture of rapidity gap formation, the determination
of the color singlet exchange fraction $f_s$ is essentially nonperturbative.
We start from a basic feature of the two-gluon exchange model: unitarity 
fixes the imaginary part of the $t$-channel two-gluon exchange amplitude in 
terms of the Born amplitude and this imaginary part dominates $t$-channel 
color singlet exchange~\cite{bjgap}. Rewriting this
relationship in terms of phase shifts, the one- and two-gluon exchange 
amplitudes are found to be too large to be compatible with
unitarity. Phase shift unitarization leads to a more realistic description, 
in which the total differential cross section remains unchanged compared 
to the Born result, but with $t$-channel color singlet exchange fractions 
which differ substantially from the expectations of the two-gluon exchange 
model. These features are demonstrated analytically for fixed values of the 
strong coupling constant, $\alpha_s$, in Section~\ref{sec2}. 
In Section~\ref{sec3} we then perform a numerical analysis for running 
$\alpha_s$, showing that the key properties of the fixed-$\alpha_s$ 
results remain unchanged. 

The predicted color singlet fractions
are found to very strongly depend on the regularization of gluon exchange at
small momentum transfer, however, and thus cannot be reliably calculated 
within perturbation theory. Within our unitarized model the non-perturbative
effects can be summarized in terms of two parameters, 
the survival probability of gaps, $P_s$, and a universal Coulomb
phase shift, $\psi_0$. Implications for the formation of gaps at the 
Tevatron are analyzed in Section~\ref{sec4}. In particular we
calculate how the gap fraction between two hard jets varies with 
jet transverse energies and jet  pseudorapidity separation and then compare 
predicted fractions with Tevatron data~\cite{D0gapnew,CDFgapnew}. 
Our conclusions are given in Section~\ref{sec5}. 

\section{Elastic scattering amplitude and unitarization}
\label{sec2}

Consider the elastic scattering of two arbitrary partons, $p$ and $P$,
\bq
p(i_1)+P(j_1)\to p(i_2)+P(j_2)\; ,
\eq
at momentum transfer $Q^2 = -t$. Here $i_1,\;\dots,\;j_2$ denote the
colors of the initial and final state partons.
The cross section and the partial wave amplitudes are completely
dominated by the forward region, $Q^2\ll s$, where the Rutherford
scattering amplitude,
\bq
\label{eq:Mborn}
{\cal M} = -8\pi\alpha_s\; {s\over t}\; T^a\otimes T'^a
= 8\pi\alpha_s\; {s\over Q^2}\; F_c = {\cal M}_0\;F_c \;,
\eq
provides an excellent approximation. Note that helicity is conserved in
forward scattering, hence spin need not be considered in the following.
The only process dependence arises from the color factor
$F_c = T^a\otimes T'^a$.

\subsection{Diagonalization in Impact Parameter and Color Space}

In order to study unitarity constraints, we need to diagonalize the amplitude
in both momentum/coordinate space and in color space. The first step is most
easily achieved by transforming to impact parameter space,
\bq
\label{eq:fourier}
T({\bf b}) = \int {d^2{\bf q}\over (2\pi)^2}\; {\cal M}({\bf q})
e^{-i{\bf q}\cdot{\bf b}}\; .
\eq
Neglecting multi-parton production processes, {\it i.e.} inelastic channels, 
unitarity of the $S$-matrix implies the relation
\bq
\label{eq:unitarity1}
\Im T({\bf b}) = {\textstyle {1\over4s}} |T({\bf b})|^2\; ,
\eq
for the full $2\to 2$ scattering amplitude $T({\bf b})$.
Eq.~(\ref{eq:unitarity1}) represents a matrix relation in color space.
More fully it can be written as
\bq
\Im T({\bf b})_{i_2 j_2, i_1 j_1} = {\textstyle {1\over4s}} \sum_{i,j}
T({\bf b})_{i_2 j_2, i j}\;  T^\dagger({\bf b})_{i j, i_1 j_1}  \;,
\label{eq:unitarity2}
\eq
where the sum runs over the dimension of the color space, $d_C=9$ for $Qq$
and $Q\bar q$ scattering and $d_C=24$ (64) for $qg$ ($gg$) elastic scattering.

Since the color factors can be written as hermitian matrices, the right-hand
side of Eq.~(\ref{eq:unitarity2}) represents a simple matrix product of the
color matrices. This product is easily diagonalized by decomposing the
color factors $F_c$ into a linear combination of projection operators onto
the irreducible color representations which are accessible in the $s$-channel,
\bq
\label{eq:col.factor}
F_c = (F_c)_{i_2j_2,i_1j_1} = \sum_k f_k\; (P_k)_{i_2j_2,i_1j_1}
                            = \sum_k f_k\; P_k \; .
\eq
For the case of quark-antiquark elastic scattering, for example, with
color decomposition $3\otimes\bar 3 = 1\oplus 8$, the color factor can be
written in terms of Gell-Mann matrices as
\bq
(F_c)_{i_2j_2,i_1j_1} = \left({\lambda^a\over 2}\right) {}_{i_2i_1}
\left({\lambda^a\over 2}\right) {}_{j_1j_2}
={4\over 9}\delta_{j_1i_1}\delta_{i_2j_2}-
{1\over 3}\left({\lambda^a\over 2}\right) {}_{j_1i_1}
\left({\lambda^a\over 2}\right) {}_{i_2j_2}
= {4\over 3}\;P_1 - {1\over 6}\; P_8\; .
\eq
For all cases, $Q\bar q$, $Qq$, $qg$ and $gg$ elastic scattering, the 
decomposition into $s$-channel projectors is summarized in 
Table~\ref{tab:colorop}. This color decomposition, combined with the 
transformation to impact parameter space, diagonalizes the unitarity 
relation for elastic scattering amplitudes.

\begin{table}[t]
\begin{center}
\caption{Representations and color operators for QCD elastic scattering.
The indices of the projection operators $P_k$ in the last column represent 
the dimensionalities of the irreducible color representations in the 
$s$-channel. Results for the $8\otimes 8$ decomposition are taken from
Ref.~\protect\cite{ddz}. }
\label{tab:colorop}
\begin{tabular}{|c|l|l|c|}
process & \multicolumn{2}{l|}{product representation and decomposition}
& color operator $F_c$ \\
\hline
&&&\\
$Qq$ ($\bar{Q}\bar{q}$) & $3\otimes 3$ & $\overline{3}\oplus 6$ &
$-\frac{2}{3}P_3+\frac{1}{3}P_6$ \\
&&&\\
$Q\bar{q}$ & $3\otimes\overline{3}$ & $1\oplus 8$ &
$\frac{4}{3}P_1-\frac{1}{6}P_8$ \\
&&&\\
$gq$ ($g\bar{q})$ & $8\otimes 3$ & $3\oplus\overline{6}\oplus 15$ &
$\frac{3}{2}P_3+\frac{1}{2}P_6-\frac{1}{2}P_{15}$ \\
&&&\\
$gg$ &  $8\otimes 8$ &
$1\oplus 8^S\!\oplus 8^A\!\oplus 10\oplus\overline{10}\oplus 27$ &
$3P_1+\frac{3}{2}P^S_8+\frac{3}{2}P^A_8-P_{27}$\\
&&&\\
\end{tabular}
\end{center}
\end{table}

\subsection{Phase Shift Analysis at Leading Order in $\alpha_s$}

Expanding the full $2\to 2$ amplitude $T({\bf b})$ into $s$-channel projectors,
$T({\bf b}) = \sum_k T_k({\bf b})\;P_k$, the individual coefficients $T_k$
are seen to satisfy the unitarity relation (\ref{eq:unitarity1}). The full 
scattering amplitude can thus be written in terms of real phase shifts 
$\delta_k({\bf b})$ for specific color representations in the $s$-channel, 
\bq
\label{eq:amp.full.impact}
T({\bf b})= \sum_k T_k({\bf b}) P_k = 
- \sum_k 2is \left(e^{2i\delta_k({\bf b})} -1 \right) P_k \; .
\eq
Within perturbation theory, the individual phase shifts $\delta_k({\bf b})$
can be expanded in a power series in $\alpha_s$. The lowest order term is
fixed by the Fourier transform of the Born amplitude (\ref{eq:Mborn}), which,
however, 
diverges at small $|{\bf q}|$ and needs to be regularized. This is 
most easily done by an infrared cutoff, $|{\bf q}|>1/R$, of the 
integral~(\ref{eq:fourier}). One can interpret this infrared cutoff as a 
consequence of confinement; the color singlet nature of hadrons at scales
larger than $\approx 1$~fm does not allow long wave-length gluons to couple
and, hence, soft gluon exchange must be suppressed. The cutoff $R$ is related
to the size of the hadronic wave-function~\cite{pumplin} and can be
considered as a nonperturbative parameter in the following.

With the cutoff $|{\bf q}|>1/R$, the Fourier transform of the Rutherford 
amplitude ${\cal M}_0$ to impact parameter space is given by
\bq
\label{eq:T0ofb}
T_0({\bf b})\;\; = \;\; 4s\; {\alpha_s\over 2}
\left(\log{R^2\over {\bf b}^2}+2(\log 2 -\gamma)\right) F_c \;\; \equiv\;\;
4s\;\delta_0({\bf b})\; F_c \; .
\eq
Here $\gamma = 0.577215\dots$ is Euler's constant, and terms of order
${\bf b}^2/R^2$ are neglected. Comparison with 
(\ref{eq:amp.full.impact}) yields
\bq
\label{eq:deltak.born}
\delta_k({\bf b}) = f_k\, \delta_0({\bf b})\; + {\cal O}(\alpha_s^2)\; ,
\eq
where the $f_k$ are taken from Table~\ref{tab:colorop}. Keeping the lowest 
order term in (\ref{eq:deltak.born}) only, the transformation back to 
momentum space can be performed analytically for the full amplitude in 
(\ref{eq:amp.full.impact}), with the result
\bq
{\cal M}({\bf q})= 8\pi\alpha_s\frac{s}{{\bf q}^2}\sum_k f_k P_k
	\exp\left(i\,\alpha_s f_k \log R^2 {\bf q}^2 + 
        {\cal O}((\alpha_s f_k)^3)\right) \; .
\label{eq:ip_sol}
\eq

As in the analogous QED case~\cite{QED}, 
the coefficient ${\cal M}_k$ of each projector 
$P_k$ is just the Born amplitude, multiplied by an infrared divergent phase 
factor {\it i.e.}
\bq
\label{eq:Munitarized}
{\cal M}  = \sum_k {\cal M}_k P_k = \sum_k f_k{\cal M}_0 e^{if_k\psi} P_k\; ,
\eq
with 
\bq
\label{eq:psiborn}
\psi = \alpha_s \log R^2 {\bf q}^2 +  {\cal O}(\alpha_s^3)\; .
\eq
This general structure has important consequences for the total differential
cross section, summed over all colors, and for the $t$-channel color singlet 
exchange rate. From (\ref{eq:Munitarized}) the color summed amplitude squared
is given by
\bq
\sum_{\rm colors}|{\cal M}|^2 = \sum_k |{\cal M}_k|^2d_k
\eq
where $d_k$ is the dimensionality of the $k$th irreducible color 
representation in the $s$-channel (see Table~\ref{tab:colorop}). Since 
each ${\cal M}_k$ equals its tree level value $f_k{\cal M}_0$, up to a 
phase, the total differential cross section remains unchanged by our 
phase shift unitarization\footnote{This is not necessarily true for other 
unitarization prescriptions. For example, for a ``$K$-matrix unitarization'', 
with $T_k({\bf b}) = 4sf_kT_0({\bf b})/(4s-if_kT_0({\bf b}))$, we numerically
find substantially reduced unitarized cross sections. For $gg$ scattering,
$d\hat\sigma(gg\to gg)/d\cos\hat\theta$
can be a factor 5 lower than the Born result, even for large scattering
angles. Such a unitarization procedure would be completely unacceptable 
phenomenologically. 
}. 

\subsection{Color Singlet Exchange Fraction}

In order to understand the rapidity gap rate in hadronic collisions, we need
the fraction of dijet events which are produced without color exchange in the 
$t$-channel. The color factor describing this situation is given by
\bq
F_s = \delta_{i_2i_1}\delta_{j_2j_1} = \openone \;,
\eq
which is just the unit operator, as far as the decomposition 
into $s$-channel projectors is concerned. 
For any given $2\to 2$ process we define the $t$-channel color singlet 
exchange amplitude, ${\cal M}_s$, as the coefficient of $F_s$. The exchange 
of color octet quanta in the $t$-channel or of yet higher color 
representations will be orthogonal to this term, {\it i.e.} no interference 
terms arise once the squared amplitude is summed over all colors, and this
makes the definition
of ${\cal M}_s$ unique. Decomposing the full amplitude as 
${\cal M}=\sum_k {\cal M}_k\; P_k = {\cal M}_s\;F_s + \dots$, the 
color singlet fraction, $f_s$ of Eq.~(\ref{eq:fBjorken}), for any
particular $2\to 2$ process, is then given by
\bq
\label{eq:18}
f_s = {\sum_{\rm colors} |{\cal M}_s\openone |^2 \over
  \sum_{\rm colors} |{\cal M}|^2 }
= {|\sum_k {\cal M}_k\; d_k |^2 \over d_C\; \sum_k |{\cal M}_k|^2d_k }\; ,
\eq
where $d_C=\sum_k d_k$ is the dimensionality of the full color space.

\begin{figure}[thb]
\epsfxsize=5.0in
\epsfysize=4.0in
\begin{center}
\hspace*{0in}
\epsffile{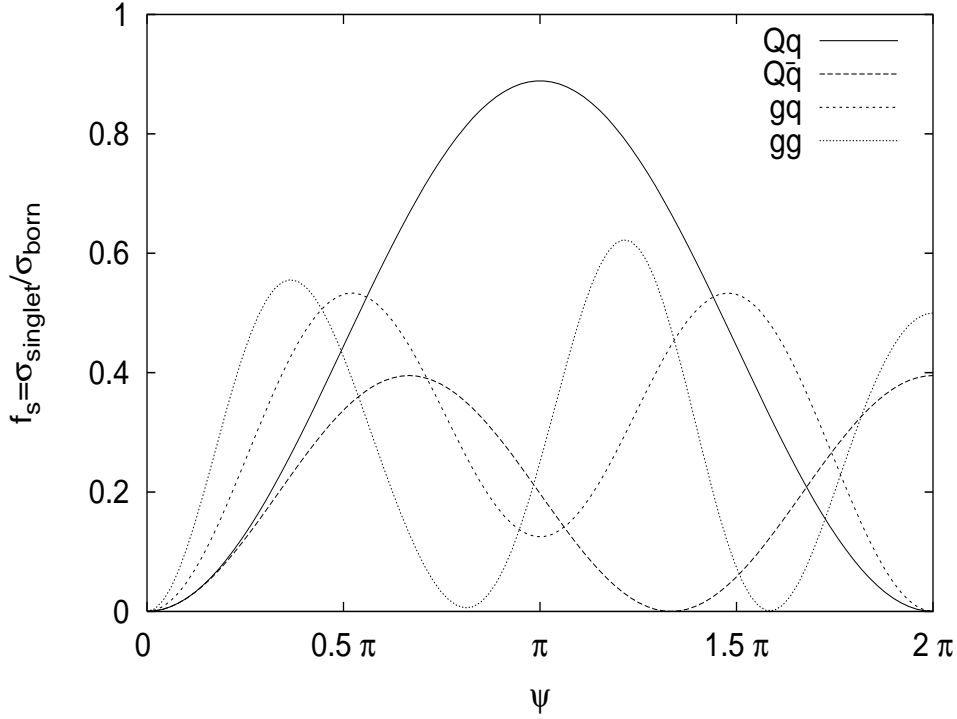}
\vspace*{-0.1in}
\caption{
Fraction of $t$-channel color singlet exchange events 
in $Qq$, $Q\bar q$, $qg$ and $gg$ elastic scattering as a function of 
the universal phase $\psi$. See text for details.
\label{fig:fs.vs.psi}
}
\vspace*{-0.1in}
\end{center}
\end{figure}

Let us apply this expression to the full, unitarized amplitude 
of Eq.~(\ref{eq:Munitarized}). The lowest order Rutherford amplitude 
${\cal M}_0$ cancels in the ratio of Eq.~(\ref{eq:18}), which hence can be 
evaluated in terms of the coefficients $f_k$ and the
dimensionalities $d_k$ of Table~\ref{tab:colorop}. We find
\ba
\label{eq:f3x3}
f_s(Qq\to Qq) \; & = & \; {8\over 9} \sin^2{\psi\over 2}\; , \\
\label{eq:f3x3bar}
f_s(Q\bar q\to Q\bar q) \; & = & \; {32\over 81} \sin^2{3\psi\over 4}\; , \\
\label{eq:f3x8}
f_s(qg\to qg) \; & = & \; {1\over 8}  \sin^2{\psi\over 2} + 
                         {15\over 32} \sin^2\psi\; , \\
\label{eq:f8x8}
f_s(gg\to gg) \; & = & \; {9\over 16} \sin^2{5\psi\over 4}\; -
                          {1\over 16} \sin^2{3\psi\over 4}\; +
                          {9\over 128}\sin^2 2\psi\; .
\ea
These color singlet fractions are plotted as a function of $\psi$ in
Fig.~\ref{fig:fs.vs.psi}.

{}From (\ref{eq:psiborn}) we see that, formally, the universal phase 
$\psi$ is of order $\alpha_s$ and thus the color singlet fractions 
are of ${\cal O}(\alpha_s^2)$, which agrees with Bjorken's result of 
Eq.~(\ref{eq:fBjorken}). Expanding (\ref{eq:f3x3bar}) to lowest order,
one obtains
\bq
f_s(Q\bar q\to Q\bar q)\; \approx\; {2\over 9}\;\psi^2 = 
{2\over 9}\;|\alpha_s \log R^2 {\bf q}^2|^2\; ,
\eq 
which appears to be four times larger than the result given 
in (\ref{eq:fBjorken}). This conundrum is resolved by observing that 
(\ref{eq:fBjorken}) represents a cross section ratio in impact parameter
space while (\ref{eq:f3x3}-\ref{eq:f8x8}) are the color singlet fractions
in momentum space. Indeed, in impact parameter space and to leading order 
in $\alpha_s$, Eqs.~(\ref{eq:amp.full.impact}-\ref{eq:deltak.born}) imply
\bq
f_s^{impact} (Q\bar q\to Q\bar q)\; \approx\; 
{2\over 9}\;|\delta_0({\bf b})|^2 
= {2\over 9}\;\left|{\alpha_s\over 2} \log {R^2\over {\bf b}^2}\right|^2\; ,
\eq 
which agrees with Eq.~(\ref{eq:fBjorken}).\footnote{For $Q\bar q\to Q\bar q$
scattering and in impact parameter space 
we thus agree with Ref.~\cite{bjgap} while the additional factor four in 
momentum space was missed in Ref.~\cite{rio}.} 
The factor four difference 
between the color singlet fractions in momentum and impact parameter space 
can be traced to a binomial factor 2 in the Fourier transform to momentum
space of 
$\log^2 R^2/b^2 = \log^2(qR)^2 - 2\log(bq)^2\;\log(qR)^2 + \log^2 (bq)^2$ 
compared to the transform of $\log R^2/b^2=\log(qR)^2-\log(bq)^2$: the 
relevant term is the one linear in $\log(bq)^2$, which is enhanced by a 
factor 2 in the first case.
Since all experiments are analyzed in momentum space, the predicted
color singlet fractions in the two-gluon exchange model need to be multiplied 
by a factor 4 compared to the results derived from (\ref{eq:fBjorken}).
For gluon-gluon scattering in particular, this would lead to a color singlet
fraction of $f_s(gg) = 0.15 \times 4 \times (9/4)^2 \approx 3$, which 
obviously cannot hold. 

The reason for this problem is the fact that, with the same arguments as used 
in Eq.~(\ref{eq:fBjorken}), the phase appearing in 
Eqs.~(\ref{eq:f3x3}-\ref{eq:f8x8}) is approximately
given by $\psi\approx 12\pi/(33-2n_f)=1.64$ which is too large to make use 
of a small angle expansion. Instead, (\ref{eq:f8x8})
predicts $f_s(gg\to gg)= 0.39$, which is still a surprisingly large 
color singlet fraction but almost an order of magnitude smaller than
the two-gluon exchange result. 

Indeed, a resummation of higher order terms is required by unitarity. 
In the two-gluon exchange model, in impact parameter space, the amplitude
for a particular color representation in the $s$-channel is given by
\bq
{1\over 4s} T_k({\bf b}) =  \delta_0({\bf b})f_k\; + \;
i(\delta_0({\bf b})f_k)^2\;
\eq
with $\delta_0({\bf b})\approx 6\pi/(33-2n_f)=0.82$, for small impact 
parameters. Here box corrections to the real part are neglected.
The unitarity relation (\ref{eq:unitarity1}), on the other 
hand, implies
\bq
\left|\Re \left({1\over 4s} T_k({\bf b}) \right)\right| < 0.5\; ,
\eq
a condition which is violated for all color channels with 
$|f_k|\ge 2/3$ in Table~\ref{tab:colorop}. Since the two-gluon exchange
model violates unitarity, we study implications of the unitarized
extension of this model in the following.

So far we have estimated the value of the universal phase $\psi$ by 
replacing $\alpha_s$ by its running value $\alpha_s(Q^2)$ 
in (\ref{eq:psiborn}),
\bq 
\psi(Q^2) = \alpha_s(Q^2) \log R^2Q^2 = {4\pi\over 11-{2\over 3}n_f}
\left(1+{\log R^2\Lambda^2\over \log Q^2/\Lambda^2}\right)
= {4\pi\over\beta_0} + \alpha_s(Q^2) \log R^2\Lambda^2 \; ,
\label{eq:psiborn.exp}
\eq
and then using the asymptotic expression for $\log Q^2\to\infty$, 
{\it i.e.} neglecting the ${\cal O}(\alpha_s)$ term in 
(\ref{eq:psiborn.exp}). This corresponds to setting the cutoff 
parameter $R=1/\Lambda$. In a more complete calculation $R$ describes 
the transverse length scale at which color screening, due to other 
partons in the proton, sets in, thus suppressing the effective gluon 
coupling. In effect, $\kappa\equiv R\Lambda\approx 1$ is a 
non-perturbative parameter for which we only have a rough guess, and which 
may be uncertain to at least a factor three, if not an order of magnitude.
With $4\pi/\beta_0= 0.52\pi=1.64$, and $\alpha_s(Q^2)\approx 0.14$ 
at momentum transfers relevant for the Tevatron rapidity gap data, a 
variation of $\kappa$ by a factor 10 leads to changes in $\psi$ by 30\% 
or more. A change of this order, in particular an increase of $\psi$, 
can drastically change the predicted color singlet fractions for individual 
processes, as is obvious from Fig.~\ref{fig:fs.vs.psi}.

\section{Running coupling effects}
\label{sec3}

An increase of $\psi$ is, in fact, to be expected when including 
running coupling effects in the determination of the tree level phase shift
$\delta_0({\bf b})$ in Eq.~(\ref{eq:T0ofb}). 
A running coupling increases the average
size of the Born amplitude (\ref{eq:Mborn}) in the Fourier transform to
impact parameter space, which leads to larger values of $\delta_0({\bf b})$ 
in (\ref{eq:T0ofb}) and this translates into a larger phase $\psi$ 
in (\ref{eq:psiborn}). 

We have analyzed this question quantitatively by determining the partial 
wave phase shifts which correspond to the Born amplitude, with running 
$\alpha_s(Q^2)$, and then unitarizing the partial wave amplitudes as
in (\ref{eq:amp.full.impact}). In terms of the Born amplitude 
${\cal M}_0(Q^2)$ the phase shifts for fixed angular momentum $J$ 
are, to lowest order, given by
\bq
\delta_J = {1\over 32\pi}\int d\cos\theta\; P_J(\cos\theta)\; 
{\cal M}_0\left(Q^2={s\over 2}(1-\cos\theta)\right)\; .
\eq
This integral is singular at $Q^2=0$, via the $1/Q^2$ pole of the Rutherford
amplitude, and at $Q^2=\Lambda^2$, via the Landau pole of $\alpha_s(Q^2)$.
Both singularities need to be regularized, for which we introduce two 
(independent) mass parameters, $M$ and $M_\alpha$. We thus replace the Born
amplitude by 
\bq
\label{eq:def.M.Malpha}
{\cal M}_0(Q^2) = 8\pi\; \alpha_s(Q^2,M_\alpha^2)\; {s\over Q^2+M^2}
= {32\pi^2\over \beta_0} \;
  {1\over \log{Q^2+M_\alpha^2\over\Lambda^2}}\;
  {s\over Q^2+M^2}\; .
\eq
We do not expect this amplitude to correctly describe the actual QCD matrix
elements at low $Q^2$. By varying the infrared cutoff parameters $M$ and 
$M_\alpha$ we rather explore the importance of the small $Q^2$ region
and, thus, the importance of non-perturbative effects which we are unable 
to calculate. The resulting unitarized amplitudes are now given by 
\bq
\label{eq:pwe}
{\cal M}_k(\theta) = 8\pi\sum_{J=0}^\infty (2J+1)P_J(\cos\theta)
{\cal T}_{k,J}
\eq
with
\bq
\label{eq:def.TJ}
i{\cal T}_{k,J} = \exp\left(i\,\frac{4\pi}{\beta_0}f_k
\int_{-1}^1\! dx\,P_J(x)\frac{1}{(z-x)\log\frac{s}{2\Lambda^2}
(z_\alpha-x)}\right)-1\; .
\eq
Here the $P_J(x)$ are Legendre polynomials, and $z=1+2M^2/s$ and 
$z_\alpha=1+2M_\alpha^2/s$ contain the two regularization parameters.

\begin{figure}[thb]
  \centering\leavevmode
  \epsfxsize=3in\epsfysize=3in\epsffile{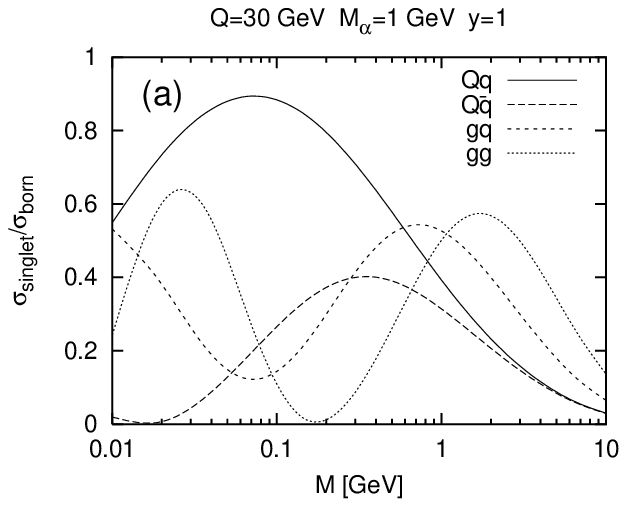}
  \epsfxsize=3in\epsfysize=3in\epsffile{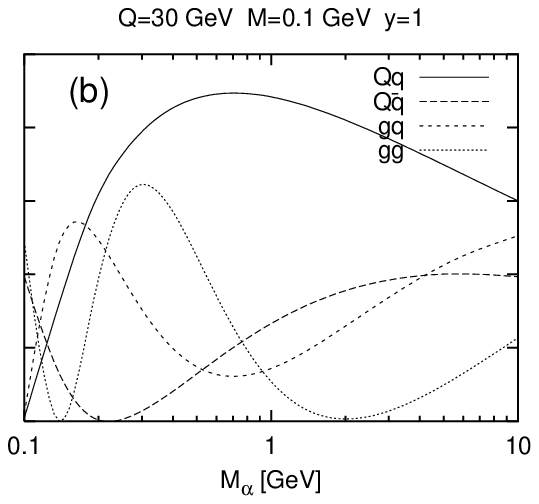}
\caption{
Dependence of the $t$-channel color singlet-exchange fraction on the
regularization parameters (a) $M$ and (b) $M_\alpha$ 
(see Eq.~(\protect\ref{eq:def.M.Malpha})). Results are shown for fixed
momentum transfer, $Q=30$~GeV, and fixed forward scattering angle.
}
\label{fig:fs.reg.dep}
\end{figure}

The integration in (\ref{eq:def.TJ}) and the partial wave sum in 
(\ref{eq:pwe}) are performed numerically.  Results are shown in
Fig.~\ref{fig:fs.reg.dep}, where possible variations of the color 
singlet fraction with the regularization parameters $M$ and $M_\alpha$ 
are explored. A very strong variation of $f_s$ is found for individual 
scattering processes. But, as we shall demonstrate below, these variations 
tend to be averaged out to a large extent when summing over the various
partonic subprocesses which contribute to actual dijet data.

The parameter dependence of $f_s$ for running strong coupling constant
in Fig.~\ref{fig:fs.reg.dep} is reminiscent of the variation of $f_s$ with
the universal phase $\psi$, which is shown in Fig.~\ref{fig:fs.vs.psi} for
our analytical results (\ref{eq:f3x3}-\ref{eq:f8x8}). The similarities
between the running coupling partial wave expansion (RPWE) and the impact 
parameter space (IP) results go far deeper, in fact~\cite{rothesis}.
Similar to the analytical IP results, the RPWE calculation leads to a
total differential cross section, summed over all colors, which
agrees with the Born result to few percent accuracy. In addition, the 
color singlet fractions at fixed $Q^2$ are found to be independent of
scattering angle or parton center of mass energy. Indeed, with 
an accuracy of a few percent, the numerical results of the RPWE calculation
can be parameterized as in (\ref{eq:f3x3}-\ref{eq:f8x8}), with a phase
$\psi = \psi(Q^2,M,M_\alpha)$ which is universal for all subprocesses. 
As is obvious from Fig.~\ref{fig:fs.reg.dep} the dependence of $\psi$ on 
the regularization parameters is quite strong. Its $Q^2$-dependence, on 
the other hand, is logarithmic only and the ansatz
\bq
\label{eq:fit.psi}
\psi(Q^2) = \psi_0+\psi_1 \log\frac{Q}{Q_0}+\psi_2 \log^2\frac{Q}{Q_0}
\eq
provides an excellent parameterization of the RPWE results. Representative 
values for the coefficients $\psi_i$ are given in Table~\ref{tab:fit.psi}. 
Only $\psi_0$ is found to depend appreciably on the regularization parameters
while $\psi_1=0.256$ and $\psi_2=-0.019$ (for $Q_0=50$~GeV)
are constant within the numerical uncertainty.
The variation of $\psi(Q)$ with momentum transfer is modest, 
with $\Delta\psi(Q) \approx \pm 0.2$ in the interval 20~GeV~$< Q < 100$~GeV.
In this region, which is of interest for the Tevatron, the RPWE values for 
$\psi(Q^2)$ in general are substantially larger than for the 
analytic impact parameter space calculation, thus confirming the qualitative
arguments made earlier.

Beyond demonstrating the relation of the numerical RPWE results to the 
analytical expressions derived in impact parameter space, the existence 
of the simple parameterization in (\ref{eq:fit.psi}) is very important 
in order to compare our calculations to experimental data. The numerical
integrations which need to be done in (\ref{eq:def.TJ}) are too slow to
be performed for individual phase space points in a Monte Carlo calculation
of dijet cross sections. With the above observations this is not necessary,
however, since instead we can use the analytical results of 
(\ref{eq:f3x3}-\ref{eq:f8x8}) together with the parameterization of
Eq.~(\ref{eq:fit.psi}).

\begin{table}[t]
\begin{center}
\caption{Dependence of the universal phase $\psi(Q^2)$ on the regularization
parameters $M$ and $M_\alpha$ (see Eq.~(\protect\ref{eq:def.M.Malpha})). 
The $\psi_i$ are the coefficients of the expansion in 
(\protect\ref{eq:fit.psi}) for $Q_0=50$~GeV.
}
\label{tab:fit.psi}
\begin{tabular}{|c|l||l|l|l|}
\multicolumn{2}{|c||}{regularization parameters } & 
\multicolumn{3}{c|}{expansion coefficients}  \\
$M$ [GeV] & $M_\alpha$ [GeV] & $\psi_0$ & $\psi_1$ & $\psi_2$ \\
\hline
0.01 & 0.2 & 8.87 & 0.254 & -0.019 \\
0.01 & 1.0 & 4.63 & 0.254 & -0.020 \\
0.01 & 5.0 & 3.25 & 0.257 & -0.017 \\
\hline
0.10 & 0.2 & 4.46 & 0.255 & -0.017 \\
0.10 & 1.0 & 3.07 & 0.255 & -0.018 \\
0.10 & 5.0 & 2.31 & 0.257 & -0.020 \\
\hline
1.00 & 0.2 & 1.71 & 0.255 & -0.020 \\
1.00 & 1.0 & 1.59 & 0.256 & -0.020 \\
1.00 & 5.0 & 1.38 & 0.257 & -0.021 \\ 
\end{tabular}
\end{center}
\end{table}

\section{Comparison with Tevatron data}
\label{sec4}

Both the D0~\cite{D0gapnew} and the CDF~\cite{CDFgapnew} collaborations at 
the Tevatron have analyzed the fraction of dijet events with rapidity gaps,
as a function of both the transverse energy, $E_T$, and the pseudorapidity
separation, $\Delta\eta=|\eta_{j_1}-\eta_{j_2}|$, of the two jets. As these 
phase space variables change, the composition of dijet events varies,
from mostly gluon initiated processes at small $E_T$ and $\Delta\eta$ 
(and, hence, small Feynman-$x$, $x_F$) to $Q\bar q$
scattering at large values. A dependence of the
gap fraction on the color structure of the scattering partons would thus be
reflected in a variation with $E_T$ and $\Delta\eta$. 

Bjorken's two gluon exchange model, which is equivalent to the small $\psi$ 
region in our analysis, predicts a larger fraction of color singlet exchange
events for gluon initiated processes~\cite{bjgap,rio} (see 
Fig.~\ref{fig:fs.vs.psi} for $\psi\alt 0.3\pi$). The gap fraction should thus 
decrease with increasing $E_T$ or $\Delta\eta$. The opposite behavior is 
expected in statistical models of color rearrangement~\cite{buchmuller,halzen}. 
Here the eight color degrees of freedom for gluons, as compared to three for 
quarks, make it less likely for gluon initiated processes that $t$-channel 
color singlet exchange is achieved by random color rearrangement. This 
would lead to a smaller gap fraction at small $x_F$ and therefore small 
$E_T$ or $\Delta\eta$. 

\begin{figure}[t]
  \centering\leavevmode
  \epsfxsize=3in\epsfysize=3in\epsffile{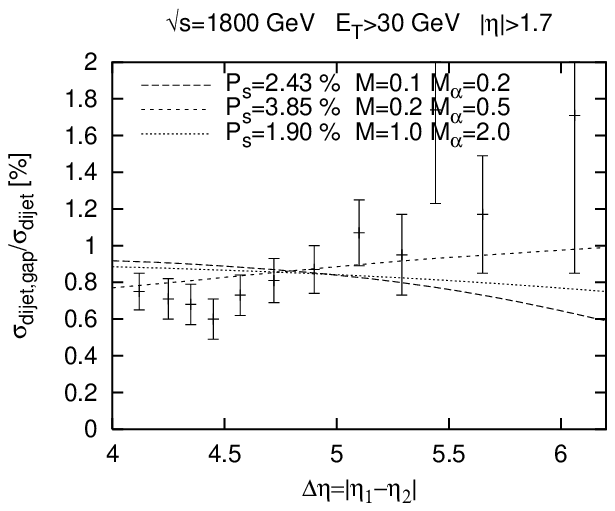}
  \epsfxsize=3in\epsfysize=3in\epsffile{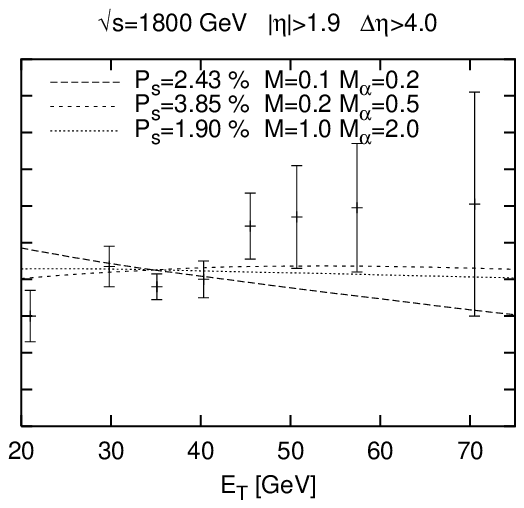}
\caption{
Dependence of rapidity gap fraction as a function of (a) $\Delta\eta$ and
(b) jet $E_T$ for three choices of regularization parameters which are given
in units of GeV. Symbols with error bars correspond 
to the D0 data as given in Ref.~\protect\cite{D0gapnew}. Also shown are 
the survival probabilities needed to reproduce the overall rapidity gap 
fraction of 0.85\% as measured by D0. }
\label{fig:fgap.exp.D0}
\end{figure}

\begin{figure}[ht]
  \centering\leavevmode
  \epsfxsize=3in\epsfysize=3in\epsffile{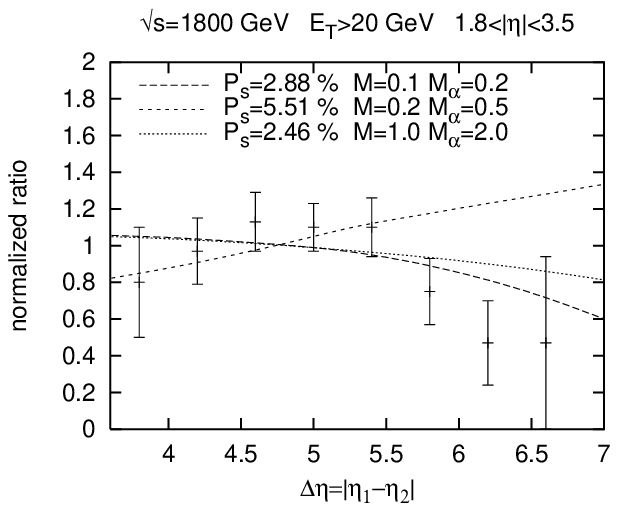}
  \epsfxsize=3in\epsfysize=3in\epsffile{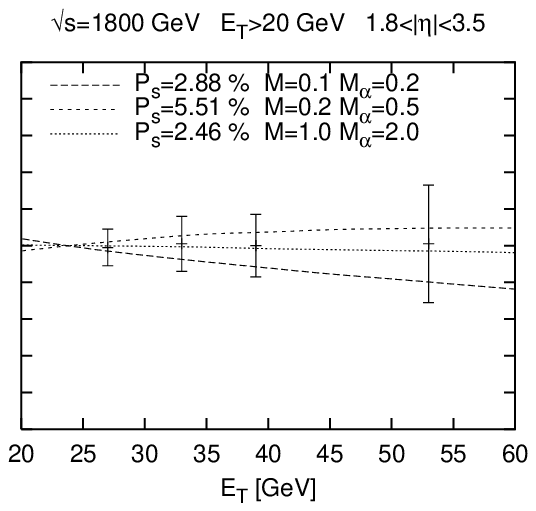}
\caption{
Dependence of rapidity gap fraction as a function of a) $\Delta\eta$ and
b) $E_T$. Shown is the gap fraction in a given bin, normalized to the 
inclusive gap fraction of 1.13\% as measured by CDF. 
 CDF data~\protect\cite{CDFgapnew} are shown with error bars together 
with predictions of the unitarized RPWE calculation, for the three choices 
of regularization parameters as in Fig.~\protect\ref{fig:fgap.exp.D0}.
}
\label{fig:fgap.exp.CDF}
\end{figure}

In the unitarized RPWE framework, the dependence on the 
regularization parameters is sufficiently strong to encompass both scenarios. 
This is demonstrated in Figs.~\ref{fig:fgap.exp.D0} and \ref{fig:fgap.exp.CDF},
where the results of the running coupling analysis for three choices of the 
regularization parameters are compared with Tevatron data, taken at 
$\sqrt{s}=1800$~GeV. The data correspond to dijet events with two opposite 
hemisphere jets of $E_T>20$~GeV, $|\eta_j|>1.8$ (CDF) or $E_T>30$~GeV, 
$|\eta_j|>1.7$ (D0). D0 data are taken from Ref.~\cite{D0gapnew} and 
show the fraction of dijet events with rapidity gaps. CDF~\cite{CDFgapnew}
shows the ratio of gap fractions in individual $\Delta\eta$ and $E_T$ bins 
to the overall gap fraction in the acceptance region. For comparing our 
calculation with the data we fix the survival probability $P_s$ in 
Eq.~(\ref{eq:Ps}) to reproduce the overall gap fraction in the acceptance 
region, which was measured as $f_{gap}=(0.85\pm 0.06\pm 0.07)\%$ for the 
D0 sample and $f_{gap}=(1.13\pm 0.12\pm 0.11)\%$ for the CDF sample. 
Required survival probabilities strongly depend on regularization parameters 
and vary between 1.9\% and 5.5\%, which is on the low side of previous 
estimates~\cite{bjgap,gotsman}. For a given choice of regularization 
parameters, predictions for the $E_T$ or $\Delta\eta$
dependence of the gap fraction are quite similar for the CDF and D0 cuts. 
To the extent that the two data sets are consistent within errors, it is not 
yet possible to discriminate between different choices of regularization 
parameters, i.e. to obtain sensitivity to the non-perturbative dynamics. 

D0 data somewhat favor color singlet fractions which grow with $x_F$ and 
which are more in line with expectations from color evaporation 
models~\cite{buchmuller,halzen}. Note that our unitarized gluon exchange 
model, with $M=0.2$~GeV and $M_\alpha=0.5$~GeV is able to describe this 
trend, even though it is an extension of the two-gluon exchange model. 
CDF data slightly prefer a gap fraction which decreases with increasing
$\Delta\eta$ and, hence, with larger $x_F$. The unitarized gluon exchange 
model, with $M=0.1$~GeV and $M_\alpha=0.2$~GeV describes such a situation.
Comparison with Fig.~\ref{fig:fs.reg.dep} shows that this set of parameters 
predicts a much smaller gap fraction for $Q\bar q$ scattering than for gluon 
initiated processes, which is qualitatively similar to the two-gluon exchange 
model~\cite{bjgap,rio}. Indeed, the shape of the gap fraction for the two 
gluon exchange model is very similar to the long-dashed curves in 
Fig.~\ref{fig:fgap.exp.CDF}. 

Clearly, the data are not yet precise enough to unambiguously distinguish 
between these different scenarios. On the theoretical side, the variation 
of the RPWE predictions with model parameters highlights the 
limitations of a perturbative approach to the color singlet exchange 
probability. Taking the phase $\psi_0$ and the survival probability $P_s$
as free parameters, the unitarized two-gluon exchange model is clearly 
capable of fitting the present Tevatron data, however.

\section{Discussion and Conclusions}
\label{sec5}

The formation of rapidity gaps in hadronic scattering events is a common 
occurrence, and its ubiquity asks for a theoretical explanation within
QCD. The formation of gaps between two hard jets at the Tevatron is 
particularly intriguing and is commonly being explained in terms of color
singlet exchange in the $t$-channel, be it via an effective color singlet
object like the ``Pomeron'' or via a statistical color rearrangement, in terms
of multiple soft gluon exchange.

``Pomeron'' exchange models build on the observation that color singlet 
exchange in the $t$-channel can be achieved in QCD via the exchange of two
gluons, with compensating colors~\cite{low}. When trying to build a 
quantitative model for the formation of rapidity gaps~\cite{bjgap}, one
encounters infrared divergences in the color singlet hard scattering 
amplitude, which in a full treatment would be 
regularized by the finite size and the color singlet nature of physical 
hadrons~\cite{pumplin}. In turn, this indicates that non-perturbative
information may be indispensable for a quantitative understanding of the 
hard color singlet exchange process. 

We have analyzed this question within a particular model, based on the 
unitarization of single gluon exchange in the $t$-channel. The Low-Nussinov 
model~\cite{low} corresponds to a truncation of the unitarization at order
$\alpha_s^2$. We find that, for any reasonable range of regularization 
parameters, the two-gluon exchange approximation violates partial wave 
unitarity, and thus a fully unitarized amplitude is needed for phenomenological 
applications.

The unitarization of hard elastic quark and gluon scattering is not unique, 
of course, but any acceptable method must preserve the successful description 
of hard dijet events by perturbative QCD. The phase shift approach used here
fulfills this requirement: the unitarization does not change the Born-level
predictions for the color averaged differential cross sections. As a corollary,
the color-inclusive dijet cross section is independent of the regularization 
parameters which need to be introduced for the full phase shift analysis.

The situation is entirely different when considering the $t$-channel color 
singlet exchange component which is introduced by the exchange of two or more 
gluons or by unitarization. The $t$-channel color singlet exchange fraction, 
$f_s$, is strongly affected 
by the full unitarization and deviates from the expectations of the
two-gluon exchange approximation, changing even the qualitative predictions
of the Low-Nussinov model. These strong unitarization effects are reflected
by a strong dependence on the precise regularization procedure. This cutoff 
dependence, which parameterizes non-perturbative effects, does not allow
to make quantitative predictions for the color singlet exchange fractions
in particular partonic subprocesses. These limitations, which have been
demonstrated here for the two-gluon exchange approximation to the Pomeron,
may be generic to Pomeron exchange models, and should be analyzed more
generally.

In spite of these limitations, we find some intriguing features of the
unitarized gluon exchange amplitudes. For all partonic subprocesses and for
all regularization parameters, the color singlet exchange fractions can 
be described in terms of a single universal phase, $\psi(Q^2)$, which
absorbs all non-perturbative effects. This suggests a unified phenomenological
description of the rapidity gap data, via the gap survival probability $P_s$
and the phase $\psi(Q^2)$. Such an analysis goes beyond the transverse 
momentum and pseudorapidity dependence of rapidity gap fractions which 
have just become available, and should best be performed directly by the 
experimental collaborations.

\section*{Acknowledgments}

We would like to thank F.~Halzen and J.~Pumplin for useful discussions 
on the physics of rapidity gaps. 
This research was supported in part by the U.S.~Department of Energy under
Grant No.~DE-FG02-95ER40896, and in part by the University
of Wisconsin Research Committee with funds granted by the Wisconsin Alumni
Research Foundation.

\end{document}